\documentclass[12pt,preprint]{aastex}

\def\lesssim{\mathrel{\hbox{\rlap{\hbox{\lower4pt\hbox{$\sim$}}}\hbox{$<$}}}}
\def\gtsim{\mathrel{\hbox{\rlap{\hbox{\lower4pt\hbox{$\sim$}}}\hbox{$>$}}}}

\def\nh3{\mbox{${\rm NH_3}$}}

\begin{document}

\title{The $c2d$ Spitzer spectroscopy survey of ices around low-mass young stellar objects, III: CH$_4$}

\author{Karin I. \"Oberg\altaffilmark{1\ast}, A. C. Adwin Boogert\altaffilmark{2}, Klaus M. Pontoppidan\altaffilmark{3}, Geoffrey A. Blake\altaffilmark{3}, Neal J. Evans\altaffilmark{4}, Fred Lahuis\altaffilmark{5} and Ewine F. van Dishoeck\altaffilmark{1}}

\altaffiltext{$^\star$}{To whom correspondence should be addressed; E-mail:  oberg@strw.leidenuniv.nl.}
\altaffiltext{1}{Leiden Observatory, Leiden University, P.O. Box 9513, NL 2300 RA Leiden, The Netherlands.}
\altaffiltext{2}{IPAC, NASA Herschel Science Center, Mail Code 100-22, California Institute of Technology, Pasadena, CA 91125, USA.}
\altaffiltext{3}{Division of Geological and Planetary Sciences, California Institute of Technology, Mail Stop 150-21, Pasadena, CA 91125, USA.}
\altaffiltext{4}{Department of Astronomy, University of Texas at Austin, 1 University Station C1400, Austin, TX 78712-0259, USA.}
\altaffiltext{5}{SRON, PO Box 800, NL 9700 AV Groningen, The Netherlands.}


\begin{abstract}
CH$_4$ is proposed to be the starting point of a rich organic chemistry. Solid CH$_4$ abundances have previously been determined mostly toward high mass star forming regions. Spitzer/IRS now provides a unique opportunity to probe solid CH$_4$ toward low mass star forming regions as well. Infrared spectra from the Spitzer Space Telescope are presented to determine the solid CH$_4$ abundance toward a large sample of low mass young stellar objects. 25 out of 52 ice sources in the $c2d$ (cores to disks) legacy have an absorption feature at 7.7 $\mu$m, attributed to the bending mode of solid CH$_4$. The solid CH$_4$ / H$_2$O abundances are 2-8\%, except for three sources with abundances as high as 11--13\%. These latter sources have relatively large uncertainties due to small total ice column densities. Toward sources with H$_2$O column densities above 2$\times 10^{18}$ cm$^{-2}$, the CH$_4$ abundances (20 out of 25) are nearly constant at 4.7$\pm$1.6\%. Correlation plots with solid H$_2$O, CH$_3$OH, CO$_2$ and CO column densities and abundances relative to H$_2$O reveal a closer relationship of solid CH$_4$ with CO$_2$ and H$_2$O than with solid CO and CH$_3$OH. The inferred solid CH$_4$ abundances are consistent with models where CH$_4$ is formed through sequential hydrogenation of C on grain surfaces. Finally the equal or higher abundances toward low mass young stellar objects compared with high mass objects and the correlation studies support this formation pathway as well, but not the two competing theories: formation from CH$_3$OH and formation in gas phase with subsequent freeze-out.
\end{abstract}

\keywords{infrared: ISM --- ISM: molecules --- ISM: abundances --- stars: formation --- astrochemistry}

\section{Introduction}

The presence and origin of complex organic molecules in protostellar regions and their possible incorporation in protoplanetary disks is an active topic of research. CH$_4$ is proposed to be a starting point of a rich chemistry, especially when UV photons are present \citep{dartois05}. In particular CH$_4$ is believed to play a key role in the formation process of prebiotic molecules \citep{markwick00}. 

CH$_4$ is less well studied in interstellar and circumstellar media compared to other small organic molecules because CH$_4$ has no permanent dipole moment and therefore cannot be observed by pure rotational transitions at radio wavelengths. Solid CH$_4$ was first detected through its bending mode at 7.67 $\mu$m from the ground by \citet{lacy91}, and with the Infrared Space Observatory Short Wavelength Spectrometer (ISO-SWS) by \citet{boogert96} toward a few high-mass sources. Tentative claims have been made toward some other objects including low-mass protostars, but are inconclusive because of the low S/N ratio of these data \citep{cernicharo00, gurtler02, alexander03}. Solid CH$_4$ has also been detected from the ground through its stretching mode at 3.3 $\mu$m, but only toward the brightest high mass sources due to problems in removing the many atmospheric lines in this spectral region \citep{boogert04}.

Models predict CH$_4$ to form rapidly on cool grains through successive hydrogenation of atomic C; similarly H$_2$O is formed through hydrogenation of atomic O \citep{vandehulst46, allen77, tielens82,brown88, hasegawa92,aikawa05}. Observations of CH$_4$ hence provide insight into the basic principles of grain surface chemistry. Compared to H$_2$O the observed gas- and solid-state CH$_4$ abundances are low; reported CH$_4$ abundances are typically a few percent with respect to H$_2$O \citep{lacy91, boogert98}. This points to relatively low atomic C abundances at the time of CH$_4$ formation, with most C already locked up in CO as H readily reacts with C on surfaces \citep{hiraoka98}. This is in agreement with the high CH$_3$OH abundances in several lines of sight, formed by hydrogenation of CO \citep{dartois99,pontoppidan03}, and large CO$_2$ abundances, formed through oxidation of CO or hydrogenated CO. That these molecules are all formed through a similar process is corroborated by the profiles of solid CO$_2$ absorption bands, which usually show an intimate mixture of CO$_2$, CH$_3$OH and H$_2$O in  interstellar ices \citep{gerakines99, boogert00, knez05}. 

If CH$_4$ is formed efficiently through grain surface reactions as well, CH$_4$ should be similarly mixed with H$_2$O. Observations of solid CH$_4$ toward a few high mass young stellar objects (YSOs) show that the CH$_4$ absorption band profiles are broad and agree better with CH$_4$ in a hydrogen bonding ice, H$_2$O or CH$_3$OH, than with a pure CH$_4$ ice or CH$_4$ mixed with CO \citep{boogert97}. This profile analysis does not, however, exclude CH$_4$ formation from photoprocessing of CH$_3$OH \citep{allamandola88, gerakines96}. In addition, because of the small sample in previous studies, it is unclear if these broad profiles are a universal feature. Hence it cannot be excluded that CH$_4$ in some environments may form in the gas phase and subsequently freeze out.

Because formation pathway efficiency depends on environment, another method for testing formation routes is through exploring the distribution of CH$_4$ toward a large sample of objects of different ages, luminosities and ice column densities. In addition, correlations, or lack thereof, with other ice constituents may provide important clues to how the molecule is formed. If CH$_4$ is formed through hydrogenation on grain surfaces in quiescent clouds, the CH$_4$ abundance with respect to H$_2$O should be fairly source independent since this mechanism mainly depends on the initial conditions of the cloud before the star forms, which seem to vary little between different star forming regions. Because of the generality of this mechanism, the CH$_4$ and H$_2$O, and possibly CO$_2$, column densities should correlate over a large range of different environments. This is also the prediction of several models where the solid CH$_4$/H$_2$O and CH$_4$/CO$_2$ ratios in dark clouds vary little both as a function of time \citep{hasegawa92} and distance into a cloud that has collapsed \citep{aikawa05}. In contrast, CO, which is formed in the gas phase and subsequently frozen out, is predicted to only correlate with CH$_4$ during certain time intervals. If CH$_4$ instead forms in the gas phase, solid CH$_4$ should be better correlated with solid CO than with solid H$_2$O and CO$_2$, since pure CH$_4$ freezes out and desorbs at similar temperatures to CO \citep[Fraser et al. submitted to $A\&A$,][]{collings04}. Finally, if CH$_4$ forms by UV photoprocessing of CH$_3$OH more CH$_4$ would also be expected to form toward sources with stronger UV-fields i.e. higher mass objects.

The objective of this study is to determine the CH$_4$ abundances and distribution pattern toward a sample of low mass young stellar objects, varying in evolutionary stage and total ice column density. The distribution pattern and correlations with other ice constituents within the sample as well as comparison with high mass young stellar objects will be used to constrain the CH$_4$ formation mechanism.  This study is based on spectra acquired with the Spitzer Infrared Spectrometer (IRS) as part of our legacy program `From molecular cores to protoplanetary disks' ($c2d$), which provides a large sample (41 sources) of infrared spectra of low mass star formation regions \citep{evans03}. In addition 11 sources are added from the GTO program 2 for which ground based 3-5 $\mu$m already exist \citep{pontoppidan03}. Overviews of the H$_2$O, CO$_2$, CH$_3$OH and other ice species in these data are found in Boogert et al. (ApJ submitted, Paper I) and Pontoppidan et al. (ApJ submitted, Paper II). The detection of solid CH$_4$ toward one of the sources, HH46 IRS, was published by \citet{boogert04b}. We have detected an absorption feature, which is attributed to solid CH$_4$, toward 25 out of 52 low mass ice sources found in this $c2d$ sample.

\section{Source sample selection, observations and data reduction}
The source sample consists of a combination of known low mass protostars and new 
protostars discovered by their Spitzer IRAC and MIPS broad-band SEDs. Spitzer/IRS spectra were obtained as part of the $c2d$ Legacy program (PIDs 172 and 179) as well as a dedicated open time program (PID 20604) and a few archival spectra observed as a part of the GTO programs of Houck et al. Among the targets observed with the IRS short-long (SL) module, 41 ice sources were identified from their spectra in the $c2d$ sample (Paper I). The GTO sources are all associated with the Ophiuchus cloud and were selected based on previous ice observations at 3-5$\mu$m \citep{pontoppidan03}. The source sample of 25 low mass protostars presented here was selected from the 41 $c2d$ and 11 GTO ice sources based solely on the existence of an absorption feature at 7.7 $\mu$m, identified with solid CH$_4$ (Table \ref{sources}). Due to the high sensitivity of the Spitzer/IRS, a large range of star formation stages are represented in the sample from very young YSOs at the interface of the Class 0/I stages (e.g. B1-c) to objects like RNO 91, identified as an inclined disk with a remnant envelope. It also includes VEry Low Luminosity Objects (VELLOs) such as L1014 IRS \citep{young04}. More information about the evolutionary stages of the objects are reported in Paper I.

The Spitzer/IRS spectra were reduced using the 2-dimensional Basic Calibrated Data (BCD) spectral images, produced by the Spitzer Science Center (SSC) pipeline version S13.2.0, as a starting point. The low resolution modules, which are the relevant ones for the solid CH$_4$ feature, were reduced in a similar way to ground-based spectra. First, the spectral and spatial dimensions were orthogonalized, and then the 2-dimensional images of the two nodding positions were subtracted in order to remove the extended emission. A fixed width extraction was performed and then the 1-dimensional spectra were averaged. Subsequently the spectra were divided by spectra of the standard star HR 2194, reduced in the same way, to correct for wavelength dependent slit losses. Finally the spectra were multiplied along the flux scale in order to match Spitzer/IRAC photometry. One source, EC 92, was carefully extracted manually due to multiple other sources present in the slit. More details about the reduction and the complete mid-infrared spectra of the $c2d$ sources is presented in Paper I.  The CH$_4$ 7.7 $\mu$m absorption feature falls within the range of the IRS modules SL1 and SL2. At these wavelengths the resolving power $R=\lambda / \Delta \lambda$ of IRS  is 65 or 0.12 $\mu$m for SL1 and 125 or 0.06 $\mu$m for SL2 (IRS Data Handbook). The plotted spectra are made up using data from the main orders of SL1 and SL2, while the SL1 bonus order was only used to confirm detections. SL2 stops around 7.6 $\mu$m so the peak of the CH$_4$ absorption feature is always in SL1. The part of the spectra from the SL2 module is hence mainly used to determine the continuum and in some cases the shape of a low wavelength wing. When comparing observations to laboratory spectra, the laboratory spectra are always convolved to the resolution of SL1 (0.12 $\mu$m, sampled with two pixels per resolution element).

ISO-SWS spectra of four high mass YSOs are used for comparison between low mass and high mass protostars in this paper. The latest Standard Processed Data (SPD) pipeline version 10.1 products were taken from the ISO archive and the detector scans were cleaned from cosmic ray hits and averaged. The final spectra do not show significant differences with respect to the data published in Keane et al. (2001). 

\clearpage
\begin{deluxetable}{lcccccc}
\tabletypesize{\scriptsize}
\tablecaption{The source sample of 25 low mass stars observed with Spitzer-IRS. In addition 4 high mass stars with previously published ISO spectra and CH$_4$ detections are included for comparison.\label{sources}}
\tablewidth{0pt}
\tablehead{
\colhead{Source} & \colhead{Alias} & \colhead{RA J2000} & \colhead{Dec J2000} & \colhead{Cloud} &
\colhead{Type} & \colhead{Obs. ID}
}
\startdata
IRAS 03235+3004&&03:26:37.5&+30:15:27.9&Perseus&low&9835520\\
IRAS 03245+3002&&03:27:39.0&+30:12:59.3&Perseus&low&6368000\\
L1455 SMM1&&03:27:43.3&+30:12:28.8&Perseus&low&15917056\\
IRAS 03254+3050&&03:28:34.2&+31:00:51.2&Perseus&low&11827200\\
B1-c&&03:33:17.9&+31:09:31:0&Perseus&low&13460480\\
B1-b&&03:33:20.3&+31:07:21.4&Perseus&low&1596544\\
L1489 IRS&IRAS 04016+2610&04:04:43.1&+26:18:56.4&Taurus&low&3528960\\
IRAS 08242-5050&HH46 IRS&08:25:43.8&-51:00:35.6&HH46&low&5638912\\
IRAS 12553-7651&&12:59:06.6&-77:07:40.0&Cha&low&9830912\\
IRAS 15398-3359&&15:43:02.3&-34:09:06.7&B228&low&5828864\\
GSS 30 IRS1&&16:26:21.4&-24:23:04.1&Ophiuchus&low&12699392\\
IRS 42&&16:27:21.5&-24:41:43.1&Ophiuchus&low&12699648\\
IRS 43&&16:27:27.0&-24:40:52.0&Ophiuchus&low&12699648\\
IRS 44&&16:27:28.1&-24:39:35.0&Ophiuchus&low&12699648\\
IRS 63&&16:31:35.7&-24:01:29.5&Ophiuchus&low&12676608\\
VSSG 17&IRS 47&16:27:30.2&-24:27:43.4&Ophiuchus&low&12698624\\
RNO 91&IRAS 16316-1540&16:34:29.3&-15:47:01.4&L43&low&5650432\\
B59 YSO5&&17:11:22.2&-27:26:02.3&B59&low&14894336\\
2MASSJ17112317-2724315&&17:11:23.1&-27:24:32.6&B59&low&14894592\\
SVS 4-5&EC 88&18:29:57.6&+01:13:00.6&Serpens&low&9407232\\
EC 92&SVS 4-10&18:29:57.9&+0.1:12:51.6&Serpens&low&9407232\\
CrA IRS 5&&19:01:48.0&-36:57:21.6&Corona Australis&low&9835264\\
CrA IRS 7 B&&19:01:56.4&-36:57:28.0&Corona Australis&low&9835008\\
CrA IRAS32&&19:02:58.7&-37:07:34.5&Corona Australis&low&9832192\\
L1014 IRS&&21:24:07.5&+49:59:09.0&L1014&low&12116736\\
\hline
W33A&&18:14:39.4&-17:52:01.3&&high&32900920\\
NGC7538 IRS9&&23:14:01.6&+61:27:20.2&&high&09801532\\
GL989&&06:41:10.1&+0.9:29:35.8&&high&71602619\\
GL7009S&&18:34:20.9&-05:59:42.2&&high&15201140\\
\enddata
\end{deluxetable}

\clearpage
\begin{deluxetable}{lcccccccc}
\tabletypesize{\scriptsize}
\tablecaption{Gaussian parameters of the observed absorption features and the CH$_4$ column densities and abundances relative to solid H$_2$O and upper limits of SO$_2$..\label{abunds}}
\tablewidth{0pt}
\tablehead{
\colhead{Source} & \colhead{$\lambda$} & \colhead{FWHM} & \colhead{$\tau _{\rm peak}$} & \colhead{$\int \tau _{7.7 \rm}$} & \colhead{$\int \tau _{\rm CH_4}$} & \colhead{N(CH$_4$)} & \colhead{N(CH$_4$)} & \colhead{N(SO$_2$)$_{\rm max}$}\\

\colhead{} & \colhead{$\mu$m} & \colhead{$\mu$m} & \colhead{} & \colhead{cm$^{-1}$} &
\colhead{cm$^{-1}$} & \colhead{10$^{17}$ cm$^{-2}$} & \multicolumn{2}{c}{ / N(H$_2$O)$\times$100}
}
\startdata
IRAS 03235+3004&7.69$\pm$0.03&0.16$\pm$0.05& 0.16$\pm$0.07&4.6$\pm$1.0&2.9$\pm$0.8&6.2$\pm$1.8&4.3$\pm$1.4&0.35$\pm$0.19\\
IRAS 03245+3002&7.59$\pm$0.03&0.20$\pm$0.08& 0.17$\pm$0.04&4.3$\pm$0.1&3.2$\pm$0.4&6.8$\pm$0.8&1.7$\pm$0.3&0.08$\pm$0.01\\
    L1455 SMM1&7.66$\pm$0.01&0.15$\pm$0.02& 0.24$\pm$0.03&6.2$\pm$0.3&4.9$\pm$0.1&10.$\pm$0.1&5.8$\pm$0.9&0.21$\pm$0.03\\
IRAS 03254+3050&7.68$\pm$0.01&0.20$\pm$0.03&0.040$\pm$0.008&1.4$\pm$0.2&0.7$\pm$0.1&1.6$\pm$0.3&4.0$\pm$0.9&0.45$\pm$0.14\\
          B1-c&7.68$\pm$0.01&0.13$\pm$0.01& 0.40$\pm$0.05&8.7$\pm$1.3&7.6$\pm$1.2& 16$\pm$3&5.4$\pm$1.4&0.11$\pm$0.03\\
          B1-b&7.70$\pm$0.01&0.10$\pm$0.01& 0.16$\pm$0.01&2.7$\pm$0.1&2.8$\pm$0.1&5.9$\pm$0.2&3.3$\pm$0.6&0.00$\pm$0.01\\
     L1489 IRS&7.68$\pm$0.01&0.13$\pm$0.01&0.037$\pm$0.006&0.93$\pm$0.04&0.64$\pm$0.04&1.4$\pm$0.1&3.1$\pm$0.2&0.14$\pm$0.01\\
IRAS 08242-5050&7.68$\pm$0.06&0.17$\pm$0.16&0.098$\pm$0.063&2.9$\pm$1.8&1.9$\pm$0.8&3.9$\pm$1.8&5.0$\pm$2.4&0.40$\pm$0.68\\
IRAS 12553-7651&7.66$\pm$0.02&0.18$\pm$0.04&0.026$\pm$0.005&0.6$\pm$0.1&0.5$\pm$0.1&1.1$\pm$0.1&3.8$\pm$0.9&0.13$\pm$0.03\\
IRAS 15398-3359&7.68$\pm$0.01&0.13$\pm$0.01& 0.22$\pm$0.03&5.1$\pm$0.6&4.2$\pm$0.8&8.8$\pm$1.7&6.0$\pm$2.0&0.20$\pm$0.06\\
        IRS 42&7.72$\pm$0.01&0.12$\pm$0.03&0.041$\pm$0.012&0.82$\pm$0.11&0.62$\pm$0.22&1.4$\pm$0.4&7.7$\pm$2.4&0.21$\pm$0.06\\
        IRS 43&7.70$\pm$0.01&0.21$\pm$0.04&0.050$\pm$0.014&1.7$\pm$0.6&0.94$\pm$0.33&2.1$\pm$0.6&6.6$\pm$2.3&0.70$\pm$0.36\\
        IRS 44&7.71$\pm$0.01&0.19$\pm$0.01&0.053$\pm$0.015&1.7$\pm$0.4&0.9$\pm$0.2&2.1$\pm$0.4&6.1$\pm$1.4&0.63$\pm$0.23\\
        IRS 63&7.70$\pm$0.14&0.14$\pm$0.13&0.042$\pm$0.026&1.0$\pm$0.4&0.72$\pm$0.31&1.6$\pm$0.7&7.9$\pm$3.8&0.42$\pm$0.59\\
        GSS 30&7.72$\pm$0.16&0.14$\pm$0.13&0.045$\pm$0.027&1.1$\pm$0.8&0.7$\pm$0.4&1.6$\pm$0.9& 11$\pm$6.2&0.60$\pm$ 0.9\\
        VSSG17&7.68$\pm$0.01&0.26$\pm$0.04&0.042$\pm$0.006&1.8$\pm$0.2&0.8$\pm$0.1&1.8$\pm$0.2& 11$\pm$2.2& 1.5$\pm$0.35\\
         RNO91&7.69$\pm$0.03&0.27$\pm$0.13&0.038$\pm$0.009&1.6$\pm$0.7&0.68$\pm$0.17&1.6$\pm$0.5&4.8$\pm$1.6&0.76$\pm$0.44\\
      B59 YSO5&7.70$\pm$0.01&0.19$\pm$0.06&0.096$\pm$0.050&3.2$\pm$3.4&1.8$\pm$0.9&3.8$\pm$2.0&2.7$\pm$1.6&0.31$\pm$0.37\\
2MASSJ17112317&7.69$\pm$0.01&0.17$\pm$0.06& 0.13$\pm$0.052&3.9$\pm$2.0&2.5$\pm$1.0&5.4$\pm$2.2&2.8$\pm$1.2&0.20$\pm$0.23\\
        SVS4-5&7.67$\pm$0.01&0.16$\pm$0.02&0.083$\pm$0.010&2.2$\pm$0.2&1.6$\pm$0.3&3.5$\pm$0.6&6.1$\pm$1.7&0.29$\pm$0.08\\
          EC92&7.71$\pm$0.01&0.07$\pm$0.10&0.066$\pm$0.032&0.67$\pm$0.21&0.88$\pm$0.47&2.0$\pm$1.2& 13$\pm$7.7&0.33$\pm$0.23\\
      CrA IRS5&7.68$\pm$0.01&0.19$\pm$0.02&0.055$\pm$0.011&1.7$\pm$0.3&1.1$\pm$0.1&2.3$\pm$0.3&6.2$\pm$1.0&0.50$\pm$0.13\\
     CrA IRS7b&7.69$\pm$0.02&0.16$\pm$0.05&0.080$\pm$0.019&2.2$\pm$0.5&1.5$\pm$0.5&3.1$\pm$1.1&3.0$\pm$1.2&0.19$\pm$0.08\\
    CrA IRAS32&7.72$\pm$0.04&0.09$\pm$0.13&0.093$\pm$0.018&1.4$\pm$0.4&1.6$\pm$0.2&3.5$\pm$0.4&6.6$\pm$2.5&0.12$\pm$0.05\\
     L1014 IRS&7.68$\pm$0.01&0.19$\pm$0.06& 0.11$\pm$0.022&3.7$\pm$1.2&2.4$\pm$0.7&5.1$\pm$1.5&7.1$\pm$2.3&0.53$\pm$0.25\\
\hline
Average&7.69&0.16&0.10&2.7&1.5&4.1&5.8&0.34\\
Standard deviation&0.03&0.05&0.09&2.0&1.4&3.5&2.7&0.35\\
\enddata
\end{deluxetable}
\clearpage

\section{Results}

\subsection{CH$_4$ column densities}
\begin{figure*}
\centering
\includegraphics[width=17cm]{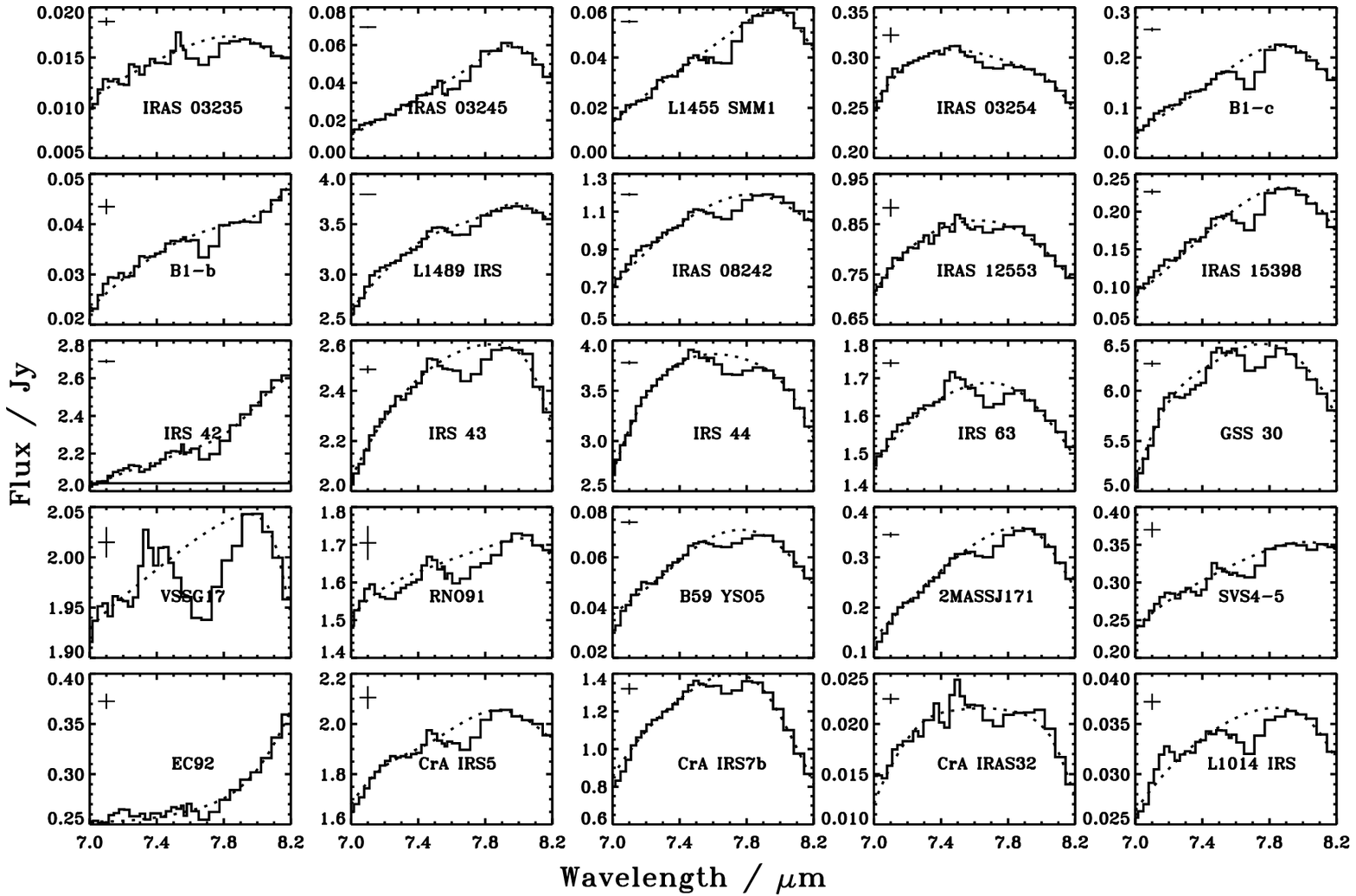}
\caption{The Spitzer-IRS spectra (solid line) of the CH$_4$ sources between 7.0 and 8.2 $\mu$m, plotted together with the chosen spline continua (dotted line). The tick bar in the upper left corner indicates the noise level derived within the $c2d$ pipeline. \label{contsp}}
\end{figure*}

Figure \ref{contsp} shows the flux calibrated spectra of the 25 low mass YSOs containing the CH$_4$ 7.7 $\mu$m absorption feature. The spectra were converted to optical depth scale using a smooth spline continuum fitted to the 7-7.4 $\mu$m and 8.0-8.3 $\mu$m regions as shown in Fig. \ref{contsp}. A local continuum is necessary since the CH$_4$ feature lies on the edge of other broader features such as the 9.7 $\mu$m feature. Fitting the continuum is complicated by the presence of a weak HI recombination emission line at 7.5 $\mu$m (Pfund $\alpha$) toward several of the sources e.g. the CrA sources, IRAS 03235+3004, L1489 IRS, RNO 91 and SVS 4-5. Of these ground based spectra around 4$\mu$m exist for CrA IRS5, CrA IRS7b, IRAS 03235+3004, L1489 IRS and SVS 4-5. In all of these spectra the 4.05 $\mu$m HI Brackett $\alpha$ is clearly visible corroborating the identification of the emission feature with hydrogen.

\begin{figure*}
\centering
\includegraphics[width=17cm]{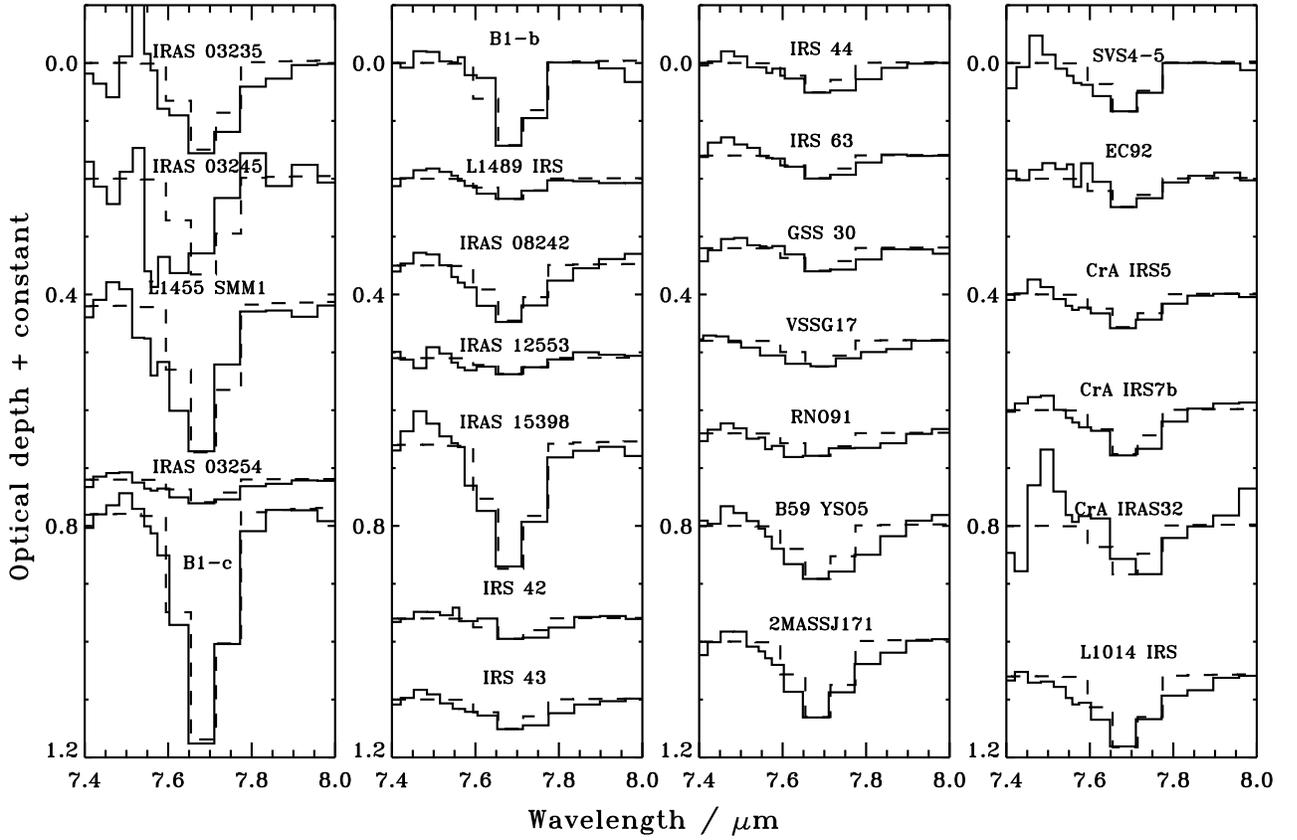}
\caption{Optical depth spectra of the CH$_4$ sources plotted together with laboratory spectra (dashed) of a CH$_4$:H$_2$O mixture 1:3 (Leiden databases at www.strw.leidenuniv.nl/$\sim$lab/databases) convolved to  at the same resolution and sampled in the same way as the observational data.}
\label{tausp}
\end{figure*}

To estimate the uncertainties introduced by the choice of continuum, the spectra were also converted to optical depth scale by adopting a local, straight line continuum between 7.4 and 7.9 $\mu$m. The differences in optical depth using the two different continua are less than 20\% for most sources, but between 30 and 60\% for a few cases: CrA IRS 7B, IRAS 03235+3004, IRAS 12553-7651 and RNO 91. The resulting optical depth spectra using the smooth spline continuum subtraction are shown in Fig. \ref{tausp}. 

The peak position, full width half maximum (FWHM) and peak optical depth of the 7.7 $\mu$m absorption feature along each line of sight were calculated from the best Gaussian fit between 7.5 and 7.9 $\mu$m (Table \ref{abunds}). The reported error estimates include uncertainties in the fit and choice of continuum. The typical peak position is 7.69 $\mu$m and the peak widths range from 0.07 to 0.27 $\mu$m. The features hence range from unresolved to barely resolved as the resolution at the center peak position is 0.12 $\mu$m. The large widths of the feature in the observed spectra, except for B1-b, suggest that CH$_4$ is generally in an H$_2$O or CH$_3$OH dominated mixture, where laboratory data show that the CH$_4$ feature has a width of up to 0.15 $\mu$m \citep{boogert97}.

\begin{figure}
\centering
\plotone{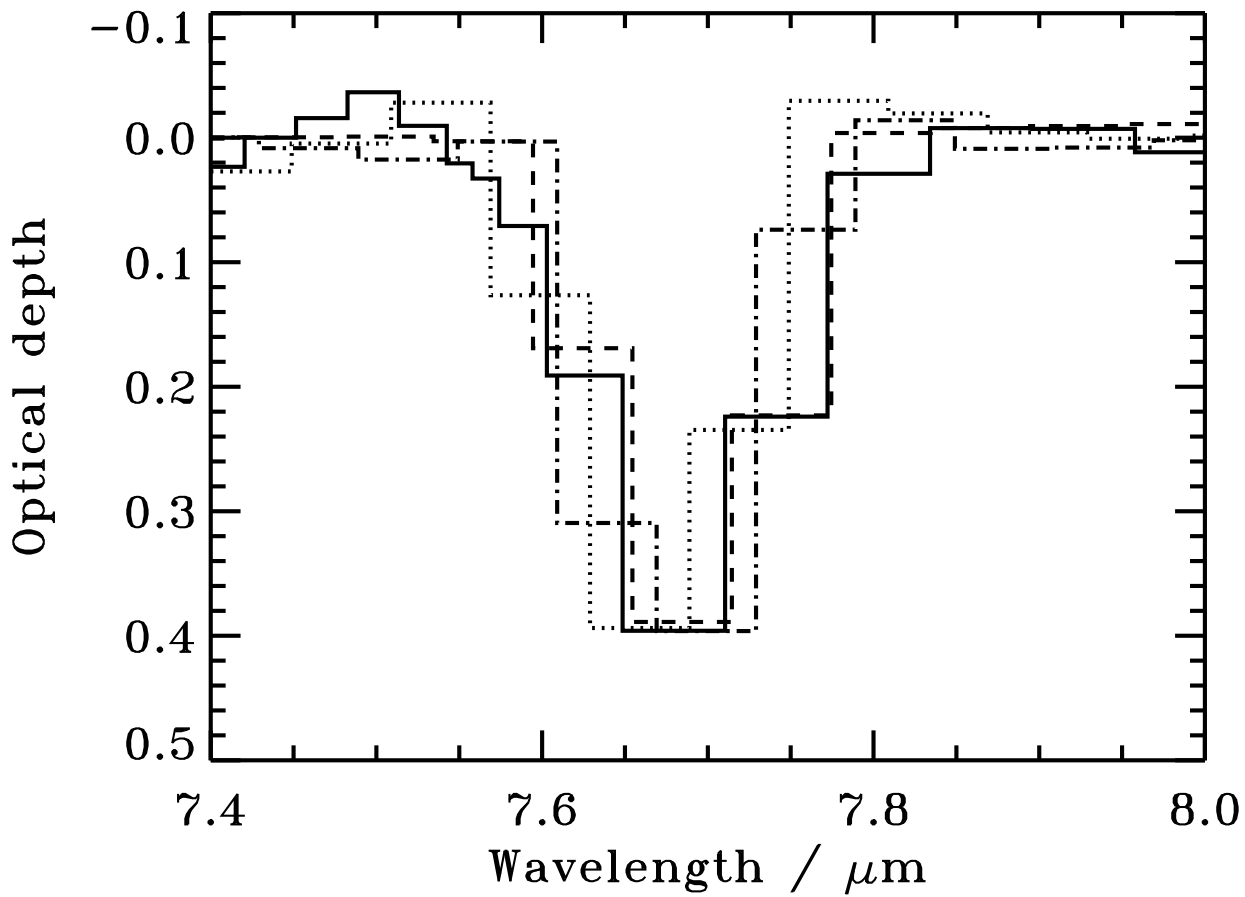}
\caption{Optical depth spectra of B1-c (solid line) plotted together with laboratory spectra of a CH$_4$:H$_2$O 1:3 mixture (dashed), a H$_2$O:CH$_3$OH:CO$_2$:CH$_4$ 0.6:0.7:1:0.1 mixture (dotted) and pure CH$_4$ ice (dashed-dotted). The derived column densities using the three different ice compositions are 1.7, 1.6 and 1.7$\times 10^{18}$ cm$^{-2}$ respectively.}
\label{mixcomp}
\end{figure}

Figure \ref{mixcomp} compares the source with the deepest absorption feature, B1-c, with different CH$_4$ containing laboratory ice spectra, convolved to the resolution of IRS-SL1. In the three laboratory ice spectra used here -- a H$_2$O dominated ice, an ice mixture that contains equal parts of H$_2$O, CH$_3$OH and CO$_2$ and pure CH$_4$, all at 10 K -- the CH$_4$ absorption profile has both different widths and peak positions. Toward most sources, as for B1-c, H$_2$O dominated ice spectra provide the best fit. Hence, comparisons with a H$_2$O:CH$_4$ 3:1 ice spectrum were used to determine the amount of CH$_4$ present in the observed spectra and also how much of the 7.7 $\mu$m feature can be accounted for by solid CH$_4$ (Fig. \ref{tausp}). Table \ref{abunds} shows that CH$_4$ can account for 45--100\% of the absorption. 

\begin{figure}
\plotone{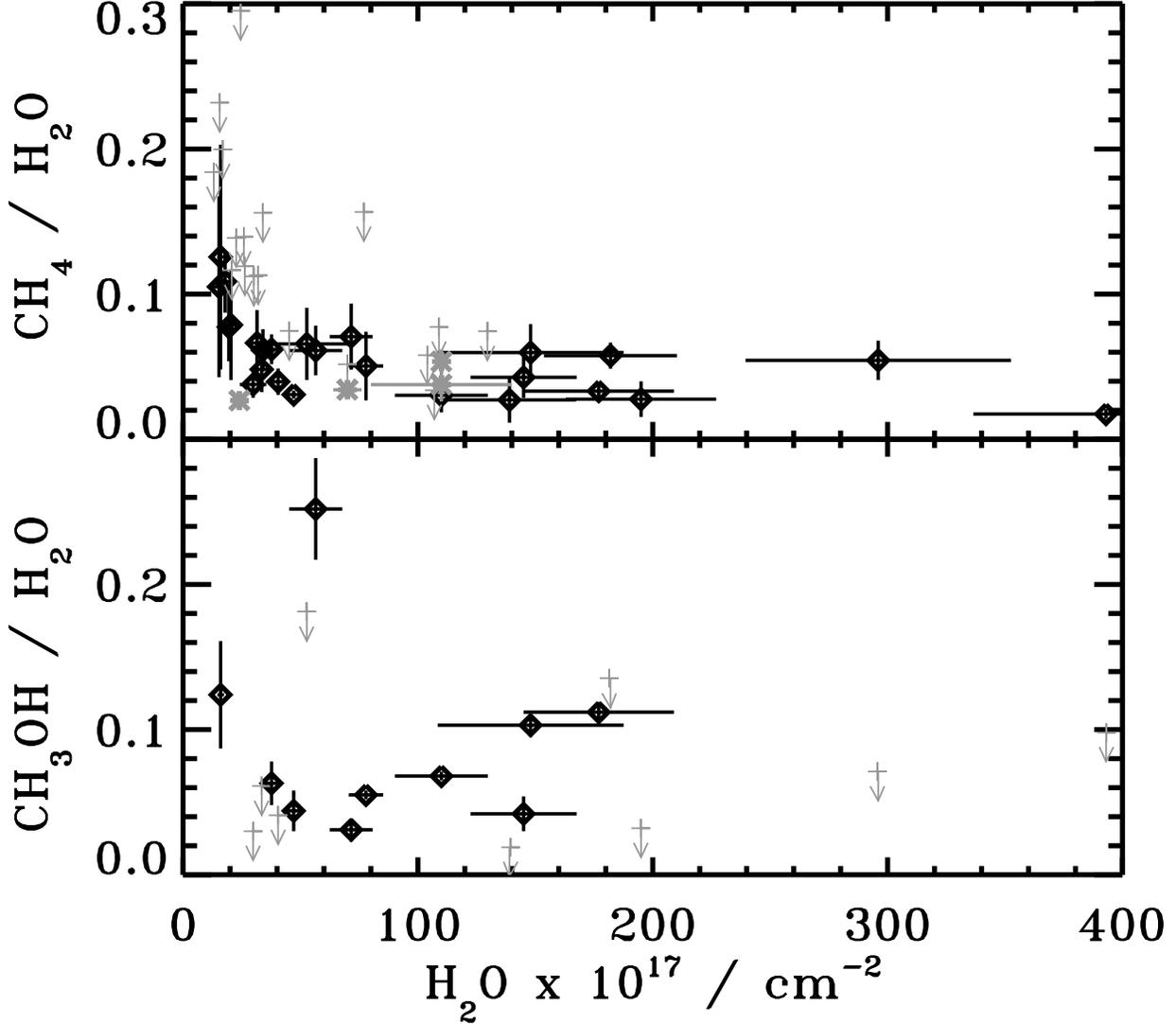}
\caption{The CH$_4$ and CH$_3$OH abundances relative to H$_2$O are plotted versus the column densities of H$_2$O for Spitzer IRS (black diamonds) and ISO (grey stars) CH$_4$ sources. Two groupings are visible in the upper plot: sources with H$_2$O column densities around 2$\times 10^{18}$ cm$^{-2}$ with high CH$_4$ abundances and sources with H$_2$O column densities  of 3--40$\times 10^{18}$ cm$^{-2}$ and a nearly constant CH$_4$ abundance around 4-5\%. CH$_3$OH abundances show no similar groupings and span a larger interval than CH$_4$. CH$_4$ and CH$_3$OH abundance upper limits below 30\% are plotted in grey.}
\label{h2och4}
\end{figure}

The CH$_4$ column densities were calculated from the integrated optical depths of the laboratory spectra, scaled to the peak optical depths of the observations, and the band strength for the bending mode of solid CH$_4$ in a H$_2$O rich ice, $4.7\times10^{-18}$ cm molecule$^{-1}$ \citep{boogert97}. The uncertainty in CH$_4$ column densities stems from both the baseline subtraction and the uncertainty in ice mixture composition even after a H$_2$O rich ice has been assumed. To obtain abundances with respect to solid H$_2$O, the CH$_4$ column densities were divided by the solid H$_2$O column densities from Paper I. Figure \ref{h2och4} shows the CH$_4$ abundances with respect to H$_2$O as a function of H$_2$O column density. The CH$_4$ abundances in the entire sample vary between 2 and 13\% and it is seen in the plot that the sample can be split into two parts: sources with H$_2$O column densities around 2$\times 10^{18}$ cm$^{-2}$ and sources with H$_2$O column densities  of 3--40$\times 10^{18}$ cm$^{-2}$. In the former group the CH$_4$ abundances with respect to H$_2$O vary between 6 and 13\% and in the latter group all CH$_4$ abundances fall between 2 and 8\%. Due to the low total column densities in the high abundance group, the uncertainties there tend to be larger. Below 3$\times 10^{18}$ cm$^{-2}$ there also seem to be some negative correlation between column density and CH$_4$ abundance. Figure \ref{h2och4} also shows the CH$_3$OH abundances and upper limits towards the sources in this sample (Paper I) . CH$_3$OH abundances span a larger interval than CH$_4$ and show none of the column density dependences visible for CH$_4$. The large variation for CH$_3$OH abundances is similar to what was found previously for a different sample by \citet{pontoppidan03}.

\begin{figure}
\centering
\plotone{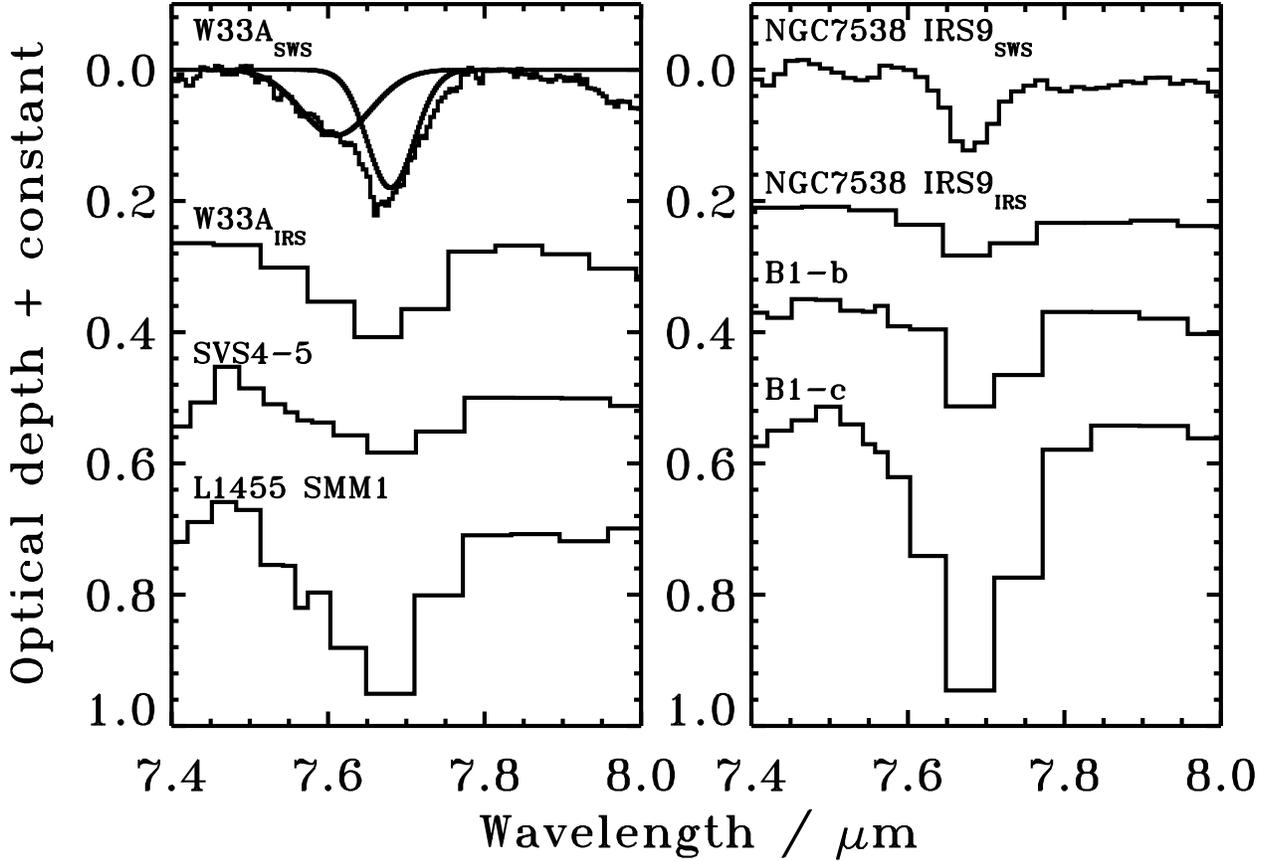}
\caption{7.7 $\mu$m profile comparison between high mass sources W33A and NGC7538 IRS9 (with their original SWS resolution and convolved to the resolution of IRS-SL1) and four low mass sources. The two Gaussians plotted together with W33A fits the observed spectrum, but only the thin component centered at 7.7 $\mu$m is explained by CH$_4$. Several of the low mass sources have similar profiles to W33A with an additional component, attributed to SO$_2$, that makes up the blue wing of the 7.7 $\mu$m feature. A few low mass sources also show thin profiles similar to that of NGC7538 IRS9.}
\label{isocomp}
\end{figure}

The excess absorption (0--55\%) of the observed 7.7 $\mu$m features in many of the astronomical objects is due to broader profiles than expected even for CH$_4$ in a H$_2$O rich ice. It is possible that an additional molecule is contributing to the optical depth of the 7.7 $\mu$m feature, e.g. solid SO$_2$. Solid SO$_2$ was suggested by \citet{boogert97} to explain the blue wing of the 7.7 $\mu$m feature in W33A. In contrast, another high mass source NGC7538 IRS9 displays no such wing. A comparison between these two sources and low mass sources from this sample shows that the same variation is present here (Fig. \ref{isocomp}) -- approximately 25\% of the sources in this study have a clear blue wing, perhaps attributable to solid SO$_2$. The maximum amounts of SO$_2$ present in the observed ices is estimated by assuming that all excess absorption is due to SO$_2$ and using its measured band strength of $\rm 3.4\times10^{-17} \: cm \: molecule^{-1}$ \citep{sandford93} (Table \ref{abunds}). The solid SO$_2$ abundances then vary between 0.1 and 1.5\% with respect to H$_2$O.

In cases where the excess is similar on both sides of the CH$_4$ absorption, another source of the large widths of the features may be the choice of continuum; the spline continuum was difficult to fit because the CH$_4$ feature is generally shallow and overlapping with other, larger, features. It is possible that the wings toward several sources are not intrinsic, but a product of this fit. This is highlighted by a comparison between Fig. \ref{contsp} and Fig \ref{tausp}, which shows that the sources with clear continua also have thinner absorption features than the average. 

\subsection{Upper limits of solid CH$_4$}

Upper limits for solid CH$_4$ were determined for the 27 sources without CH$_4$ detections, in the sample of 52 low mass ice sources originally probed for CH$_4$, by estimating the maximum amount of a H$_2$O:CH$_4$ 3:1 ice that could hide under the noise. The average 3$\sigma$ upper limit is 15\% solid CH$_4$ with respect to solid H$_2$O, which is in the upper range of CH$_4$ abundances in the sources with solid CH$_4$ detections. All abundance upper limits below 30\% are shown in Figure 4. Only one CH$_4$ upper limit, of $\sim$3\%, falls below the average abundance of 4.8\%. The lack of detection in these 27 sources is hence probably due to the spectral quality and low fluxes of the objects rather than a lower amount of solid CH$_4$. 

\subsection{Molecular correlations}

The objects in this sample may vary significantly in temperature structure as well as other environmental factors. Lack of correlations between molecular abundances may hence be due to either different formation pathways or differences in volatility. A lack of correlation between CH$_4$ and molecules of different volatility does hence not exclude that they formed in a similar manner. In pure form CH$_4$ has a similar volatility to CO \citep{collings04}. If CH$_4$ is mixed with H$_2$O significant amounts can be trapped inside of the H$_2$O ice, however, and then CH$_4$ has an effective volatility closer to that of H$_2$O and CO$_2$.

\begin{figure}
\plotone{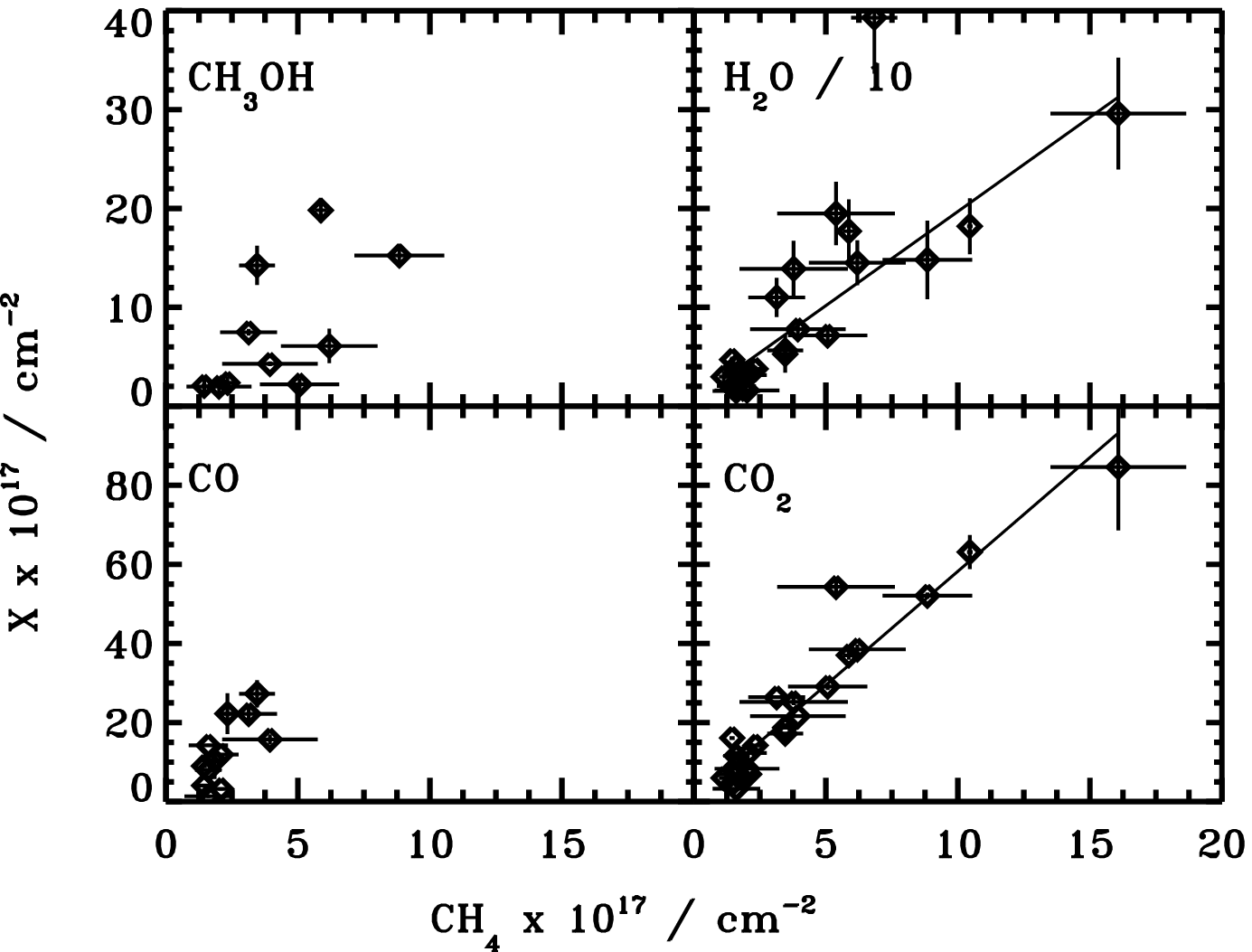}
\caption{The column densities of four ice species, CH$_3$OH, H$_2$O, CO and CO$_2$ are plotted versus the column densities of solid CH$_4$. Solid CH$_4$ is strongly correlated with solid CO$_2$ and H$_2$O. CH$_4$ is only weakly correlated with solid CO and CH$_3$OH, but this may be partly due to the fact that CO has only been observed and CH$_3$OH only detected towards much fewer targets. The outlier in the H$_2$O plot is IRAS 03245+3002.}
\label{absmol}
\end{figure}

\begin{figure}
\plotone{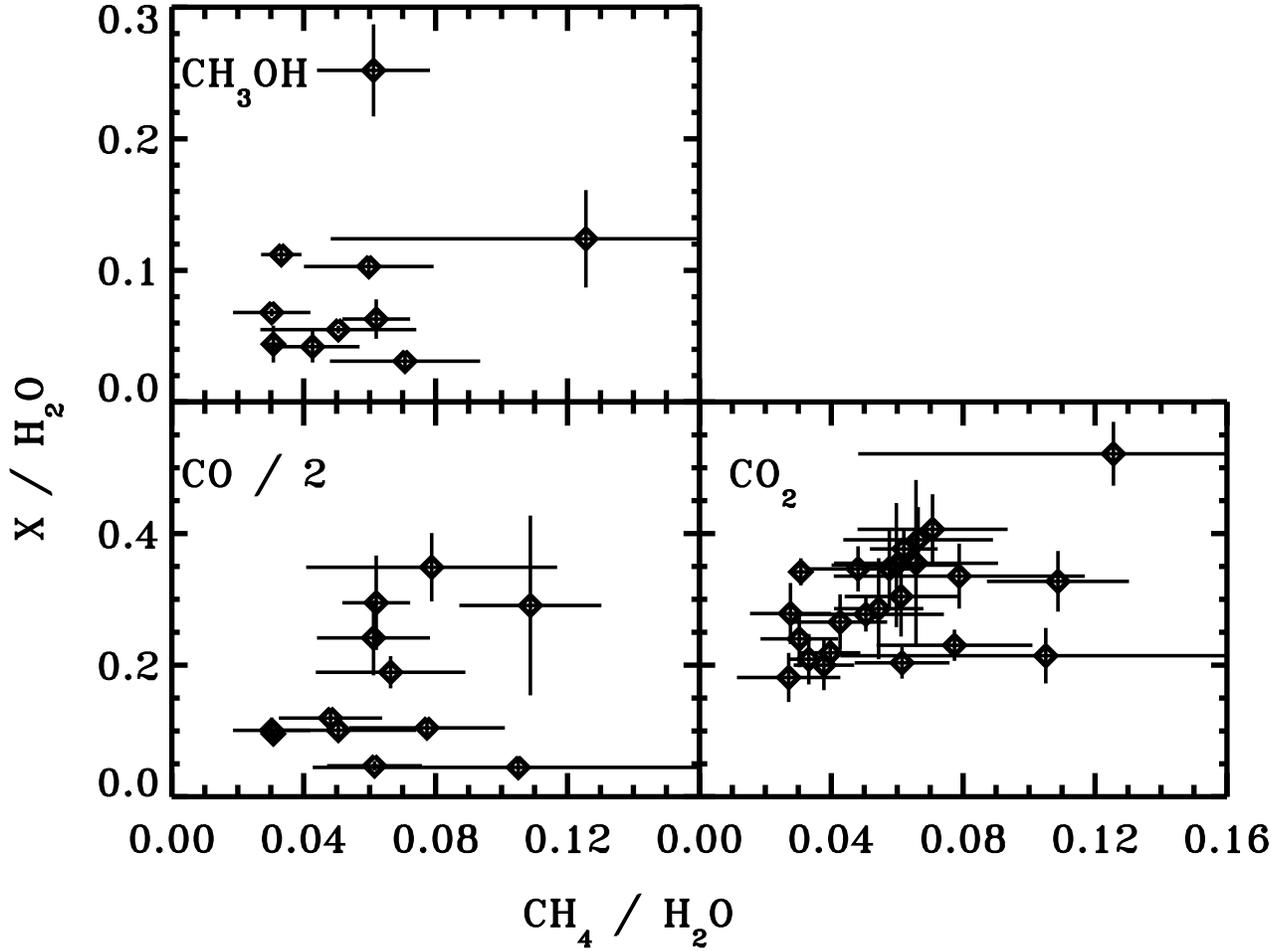}
\caption{The abundances of solid CH$_3$OH, CO and CO$_2$ relative to H$_2$O are plotted versus those of solid CH$_4$. There is no correlation between the relative amounts of solid CH$_4$ and CH$_3$OH or CO. In contrast the CH$_4$ and solid CO$_2$ abundances are weakly correlated.}
\label{h2omol}
\end{figure}

The column densities of solid H$_2$O (26 detections) and CH$_3$OH (10 detections) have been derived in Paper I and solid CO (13 ground based observations) and CO$_2$ (25 detections) in \citet{pontoppidan03} and Paper II for many of the CH$_4$ sources. The column densities and abundances for these molecules are plotted versus CH$_4$ in Figs. \ref{absmol} and \ref{h2omol}. Where the plots reveal a linear relationship between molecules the best linear fit is also drawn. The Pearson product-moment correlation coefficient, $R$, was calculated to measure the strength of the correlation -- $R^2$ directly gives the fraction of variance of the second molecule that is due to changes in CH$_4$, assuming a linear relationship between the two molecules. 

Figure \ref{absmol} shows the correlations between the column densities of solid CH$_4$  and the four other ice components. Some correlation between column densities of ice species and with total column density is always expected, but Fig. \ref{absmol} shows that the strength of this correlation is variable between CH$_4$ and the different molecules. CH$_4$ correlates strongly ($R^2$=0.91) with CO$_2$, which is believed to form on grain surfaces. The correlation with H$_2$O, another species formed on surfaces, is weaker ($R^2$=0.64), but this correlation coefficient is significantly increased if one outlier, IRAS 03245+3002,  is removed. $R^2$ is then 0.82 for the CH$_4$--H$_2$O correlation. IRAS 03245+3002 could have been identified previous to the correlation studies as an outlier since it is the only source with no CO$_2$ ice detection putting an upper limit on its CO$_2$/H$_2$O of 15\%, compared to 30$\pm$9\% for the entire sample. Even without removing the outlier the correlations are significant at the 99\% level with 23 and 24 degrees of freedom (DF) respectively.

CH$_4$ and CH$_3$OH and CO are barely signficantly correlated ($R^2$=0.48 with 8 DF and 0.49 with 11 DF respectively). The small number of CO abundances, which are only available for 11 of the targets in this study, complicates interpreting the low correlation plots between CO and CH$_4$. If only the sources for which CO measurements exist are used to calculate the CH$_4$--CO$_2$ and CH$_4$--H$_2$O correlations these are reduced to $R^2$=0.46 and 0.74 respectively. 

The stronger correlation between CH$_4$ and CO$_2$ compared to CH$_4$ and H$_2$O is curious, but may be an artefact of the fact that the CO$_2$ column densities are better known for this sample than the H$_2$O column densities. For most sources in this sample the H$_2$O column densities are determined by estimating the depth of the H$_2$O libration feature, which is difficult due to the overlap with the silicate feature at 9.7 $\mu$m. The uncertainties in these H$_2$O column densities were estimated to $\sim$30\% in Paper I after comparison between derived abundances from the 3 $\mu$m mode and the libration mode for the sources with 3 $\mu$m spectra. Furthermore the different amounts of CO$_2$ toward the different sources may introduce further error in the derived H$_2$O column densities as recently shown in \citet{oberg07}. 

Figures \ref{h2omol} show the correlations between solid CH$_4$ and the other three ice species, with the abundances normalized to the solid H$_2$O column density - the main ice constituent (Fig. \ref{h2omol}). When normalizing to H$_2$O there is no significant correlation at a 95\% level between CH$_4$ and CH$_3$OH and CO ($R^2$ below 0.1) and at best a weak correlation between CH$_4$ and CO$_2$ ($R^2$ = 0.27). There is also no correlation between CH$_4$ abundances and any of the three CO components as defined by \citet{pontoppidan03}, with the different components corresponding to pure CO and CO in a H$_2$O and CO$_2$ rich ice respectively. 

\subsection{Spatial trends}

The average CH$_4$ abundance relative to H$_2$O and its standard deviation is shown for all clouds with more than one detection and for the entire sample in Table \ref{cloudt}. The averages for the individual clouds range from 2.8\% to 9\%, with Serpens and Ophiucuhs at 9 and 8\% respectively. Whether the clouds are significantly different in their CH$_4$ abundances was evaluated using Analysis of Variance with the Statistics101 resampling software. Resampling is more robust than traditional statistical tests since there is no need to assume an underlying distribution, which is especially useful when the sample size is small. The test procedure starts with calculating the sum of the absolute deviations of the cloud averages from the sample average. The CH$_4$ abundances in the five clouds are then randomly resampled into five new groups with the same group size distribution as before and the absolute deviations of the group averages from the sample average is re-calculated. This is repeated 1000 times and the number of times the resampled sum of deviations exceeds the sum of deviations of the cloud averages is counted. The nul-hypothesis that the difference between the clouds is due to chance could then be rejected with 95\% confidence. If Ophiuchus is removed from the sample there are no longer any significant differences between the clouds.

\clearpage
\begin{deluxetable}{lccc}
\tabletypesize{\scriptsize}
\tablecaption{The average CH$_4$ abundance relative to H$_2$O, and its standard deviation for five of the investigated clouds and for the entire sample.\label{cloudt}}
\tablewidth{0pt}
\tablehead{
\colhead{Cloud} & \colhead{Number of sources } & \colhead{Average} & \colhead{St. dev.}
}
\startdata       
   Perseus&6&0.040&0.014\\
Ophiuchus&6&0.083&0.019\\
       B59&2&0.027&--\\
       CrA&3&0.052&0.019\\
   Serpens&2&0.093&--\\
\hline
      All&19&0.058&0.027\\
\enddata
\end{deluxetable}
\clearpage

\section{Discussion}

\subsection{Low vs. high mass YSOs}

For comparison the solid CH$_4$ abundances with respect to H$_2$O  for several high mass YSOs (Table \ref{sources}) were re-derived and plotted relative to H$_2$O abundances from \citet{gibb04} in Fig. \ref{h2och4}. These sources have been previously investigated for solid CH$_4$ by \citet{boogert96} and \citet{gibb04}. The solid CH$_4$ abundances derived here are 30 to 60\% higher than those published by \citet{gibb04} mainly due to a difference in band strength (7.3$\times 10^{-18}$ molecules cm$^{-1}$ by \citet{gibb04} and 4.7$\times 10^{-18}$ molecules cm$^{-1}$ here). The plot shows that the high mass ISO CH$_4$ sources fit in seamlessly with the low mass CH$_4$ sources with H$_2$O column densities above 2$\times 10^{18}$ cm$^{-2}$. Figure \ref{isocomp} also shows that there is a similar variation in the 7.7 $\mu$m feature profile between low and high mass objects. There is hence no reason to expect large systematic differences in CH$_4$ abundances or formation pathways toward low and high mass YSOs. 

\subsection{Formation scenarios}

The similar or higher abundance of CH$_4$ with respect to H$_2$O toward the low mass sources in this sample compared to what has been found previously toward high mass young stellar objects, suggests that the
CH$_4$ formation rate is not dependent on stellar UV irradiation. Of the three formation scenarios for CH$_4$ suggested in the introduction, this study hence does not support a formation pathway connected to stellar UV processing of CH$_3$OH. Furthermore, in the sources with 2--40$\times 10^{18}$ cm$^{-2}$ H$_2$O the CH$_4$ abundance is nearly constant. In comparison the CH$_3$OH abundances vary by a factor of 10 and it seems unlikely that this would be the outcome if the main formation pathway of CH$_4$ is connected to CH$_3$OH through e.g. cosmic-ray induced UV processing of CH$_3$OH, which is also present under quiescent conditions.

Of the two remaining scenarios, formation in the gas phase with subsequent freeze-out and hydrogenation of C on grain surfaces, this study lends most support to the latter mechanism. First, under quiescent conditions, gas phase models predict steady state total CH$_4$/H$_2$ abundances of only around 10$^{-7}-5\times  10^{-14}$ \citep{woodall06,bergin95}, compared with our inferred CH$_4$/H$_2$ abundances of $\sim 2-13 \times 10^{-6}$ (assuming a standard H$_2$O/H$_2$ ratio of 10$^{-4}$). In early times when C/CO$>$1, the CH$_4$/H$_2$ can reach above 10$^{-6}$ \citep{millar91}. Ices cannot form at extinctions lower than $\sim2$ A$_V$, however \citep{cuppen07}. At extinctions higher than 2 A$_V$, C/CO$<$1 and hence freeze-out of the high CH$_4$ abundances at early times cannot be used to explain the high ice abundances. In addition, pure CH$_4$ has a similar volatility to CO, within a few degrees \citep{collings04}, and if both are formed in the gas phase and subsequently frozen out, the two molecules should correlate, both in absolute column densities and in abundances relative to the much less volatile H$_2$O. This study clearly shows that this is not the case.

This weak correlation between CO and CH$_4$ can be contrasted with the stronger correlations between CH$_4$ and CO$_2$ and H$_2$O column densities. Furthermore the lack of significant correlation with any species once the correlation with H$_2$O has been effectively divided out, by normalizing with the H$_2$O column densities, shows that CH$_4$ is not significantly better related to any other molecule than to H$_2$O. Together this suggests a formation scenario of CH$_4$ more related to H$_2$O and CO$_2$ formation, which both form on grains, rather than to CO. In theory the correlations could be single-handedly due to the fact that CH$_4$ has a higher desorption temperature than in its pure form due to mixing with H$_2$O. This, however, is only expected to occur if CH$_4$ is formed together with H$_2$O and hence CH$_4$ formation on grain surfaces is still the most plausible outcome of this study. 

The hydrogenation scenario is also supported by the broad profiles of the features which have only been found in laboratory spectra when CH$_4$ is in a mixture with hydrogen bonding molecules, which are believed to form on grain surfaces. These conclusions are in agreement with studies of the solid CH$_4$ profiles toward high mass YSOs, which show that solid CH$_4$ is mixed with H$_2$O or possibly CH$_3$OH \citep{boogert97} and also with the observed low gas/solid ratio of CH$_4$ compared to CO \citep{boogert04}. 

\subsection{Differences between clouds}

Figure \ref{h2och4} and Table \ref{cloudt} shows that the CH$_4$ abundances differ from cloud to cloud. In Fig \ref{h2och4}, 4 of the 5 sources with high CH$_4$ abundances and H$_2$O column densities below 2$\times 10^{18}$ cm$^{-2}$ are Ophiuchus sources (the fifth is in Serpens). The two Ophiuchus sources that have higher H$_2$O ice abundances also have more 'normal' CH$_4$ abundances. This indicates that it is the low total column density towards four of the Ophiuchus sources rather than initial conditions in Ophiuchus as a whole that is responsible for the extreme CH$_4$ abundances found there. This may be due to that towards sources with low total column densities the C/CO ratio may be larger allowing for more CH$_4$ to form. The C/CO ratio is dependent on extinction for low extinctions, but not for high ones. This is consistent with our observation that CH$_4$ abundances are only dependent on total column densities for low column densities. Sakai et al. (ApJ in press) have instead suggested that different CH$_4$ abundances could be due to different collapse times. In clouds that collapse very fast, chemical equilibrium may not be reached favoring a C-based rather than CO-based chemistry. This question can only be settled if gas phase data for these sources become available.

\subsection{Comparison with models}

Solid CH$_4$ abundances have been modelled previously for a variety of conditions \citep{hasegawa92, aikawa05}. The predicted CH$_4$ ice abundances with respect to H$_2$O
usually vary between $\approx$1--10\%, with the main formation path being sequential hydrogenation of C atoms on grains at a time when a large fraction of the gas phase C has already been converted to CO, i.e. C/CO$<$1. The observed abundances in our sample fall mostly within this range and are hence in general agreement with the models.

\citet{aikawa05} model molecular abundances on grain surfaces and in the gas phase during the collapse of a spherical cloud. Regardless of initial conditions and the central density when the collapse is stopped, the CH$_4$/H$_2$O and CH$_4$/CO$_2$ ice ratios are fairly constant as a function of radius in the cloud. The value of CH$_4$/H$_2$O ratio varies with initial conditions between a few and 10\%. In contrast the CH$_4$/CO ratio is radius dependent. The model of \citet{hasegawa92} is more aimed at modelling interstellar clouds before collapse, but as they show the time evolution of the chemistry it may still be useful to compare with our observations. As in the collapse model, CH$_4$, CO$_2$ and H$_2$O trace each other fairly well, while CO and CH$_4$ are only correlated during some time intervals, regardless of initial conditions. Both these models are consistent with Fig. \ref{absmol} where solid CH$_4$ is better correlated with H$_2$O and CO$_2$ than with CO.

\section{Conclusions}
We present Spitzer-IRS spectra of the solid CH$_4$ feature at 7.7 $\mu$m toward a large sample of low mass young stellar objects. Our conclusions are as follows:
\begin{itemize}
\item 25 out of 52 low mass young stellar objects show a solid CH$_4$ feature at 7.7 $\mu$m.
\item The solid CH$_4$ abundance with respect to H$_2$O is centered at 5.8\% with a standard deviation of 2.7\% in the sources with CH$_4$ detections. In the sources without detections the average upper limit is 15\%, which is not significant compared with the detections.
\item The sources (two Ophiuchus and one Serpens) with more than 10\% CH$_4$ all have H$_2$O column densities below 2$\times 10^{18}$ cm$^{-2}$. Due to the low total column densities, two  of these three sources have uncertainities larger than 50\%. Above 2$\times 10^{18}$ cm$^{-2}$ the sources (20 out of 25) have a fairly constant CH$_4$ abundance of 4.7$\pm$1.6\%.
\item The 7.7$\mu$m feature profiles are signficantly broader for all but one object than expected for pure solid CH$_4$ and toward most sources also broader than expected for CH$_4$ in H$_2$O dominated ices. Approximately 30\% of the features have a blue wing, seen previously toward high mass YSOs and there attributed to solid SO$_2$
\item The column densities of solid CH$_4$ and H$_2$O and CO$_2$ are clearly correlated, while CH$_4$ and CO and CH$_3$OH are only weakly correlated.
\item There is also no correlation between the CH$_4$ and CO abundances when both have been normalized to the H$_2$O abundance.
\item The Ophiuchus cloud has significantly higher CH$_4$ abundances compared to the rest of the sample, probably due to the low total column densities towards several of the sources. There are no significant differences between the remaining clouds.
\item The abundance variation is smaller for CH$_4$ compared to solid CH$_3$OH; CH$_4$ seems to belong to the class of molecules, also including H$_2$O and CO$_2$ that appear 'quiescent', i.e. their abundances are more or less constant, in contrast to highly variable ices like CH$_3$OH and OCN$^-$. If the Ophiuchus sources are included CH$_4$ is somewhere between the two classes.
\item Sample statistics and comparison with model predictions support CH$_4$ formation through hydrogenation of C on grain surfaces.
\end{itemize}

\acknowledgements
We thank Claudia Knez and the Spitzer $c2d$ IRS team for useful comments on the manuscript. Funding for KI\"O and EFvD was provided by NOVA, the Netherlands Research School for Astronomy, a grant from the European Early Stage Training Network ( MEST-CT-2004-504604), and a NWO Spinoza grant. Support for KMP was provided by NASA through Hubble Fellowship grant 1201.01 awarded by the Space Telescope Science Institute, which is operated by the Association of Universities for Research in Astronomy, Inc., for NASA, under contract NAS 5-26555.

\clearpage
			
\clearpage

\clearpage	   

\clearpage

\clearpage

\end{document}